\documentclass{caps}
\usepackage{peter}

\usepackage[dvips]{color}
\include{colordefs}
\usepackage{graphics}

\def\iaucirc{{IAU Circ.}}
\def\aa{{A\&A}}

\def\aj{{AJ}}

\def\apj{{ApJ}}

\def\apjs{{ApJS}}
\def\baas{{BAAS}}
\def\mnras{{MNRAS}}
\def\nat{{Nature}}
\def\pasp{{PASP}}
\def\etal{\emph{et al.}\ }
\def\gtrsi{\mathrel{\hbox{\rlap{\hbox{\lower4pt\hbox{$\sim$}}}\hbox{$>$}}}}
\def\lesssi{\mathrel{\hbox{\rlap{\hbox{\lower4pt\hbox{$\sim$}}}\hbox{$<$}}}}

\begin{document}
\pagenumbering{arabic}

\author[A. V. FILIPPENKO and D. C. LEONARD]{ALEXEI V. FILIPPENKO\\
Department of Astronomy, University of California, Berkeley, CA 94720-3411
\and
DOUGLAS C. LEONARD\\
Five College Astronomy Department, University of Massachusetts, Amherst, MA 01003-9305}

\chapter{Spectropolarimetric Observations of Supernovae}

\begin{abstract}

{\it We briefly review the existing database of supernova spectropolarimetry,
concentrating on recent data and on results from our group's research.
Spectropolarimetry provides the only direct known probe of early-time supernova
geometry. To obtain reliable conclusions, however, it is very important to
correctly account for interstellar polarization. We find that Type IIn
supernovae (SNe~IIn) tend to be highly polarized, perhaps in part because of
the interaction of the ejecta with an asymmetric circumstellar medium. In
contrast, SNe~II-P are not polarized much, at least shortly after the
explosion. At later times, however, there is evidence for increasing
polarization, as one views deeper into the expanding ejecta.  Moreover,
core-collapse SNe that have lost part (SN~IIb) or all (SN~Ib) of their hydrogen
(or even helium; SN~Ic) layers prior to the explosion tend to show substantial
polarization; thus, the deeper we probe into core-collapse events, the greater
the asphericity.  There is now conclusive evidence that at least some SNe~Ia
are intrinsically polarized, although only by a small amount. Finally, SN
spectropolarimetry provides the opportunity to study the fundamental properties
of the interstellar dust in external galaxies. For example, we have found
evidence for extremely high polarization efficiency for the dust along the
line-of-sight to SN 1999gi in NGC 3184.}

\end{abstract}

\section{Introduction}

Since extragalactic supernovae (SNe) are spatially unresolvable during the very
early phases of their evolution, explosion geometry has been a difficult
question to approach observationally.  Although SNe are traditionally assumed
to be spherically symmetric, several pieces of {\it indirect} evidence have
cast doubt on this fundamental assumption in the past decade, especially for
core-collapse events.  On the observational front, high-velocity ``bullets'' of
matter in SN remnants (e.g., Taylor \etal 1993), the Galactic distribution and
high velocities of pulsars (e.g., Kaspi \& Helfand 2002, and references
therein), the aspherical morphology of many young SN remnants (Manchester 1987;
see, however, Gaensler 1998), and the asymmetric distribution of material
inferred from direct speckle imaging of young SNe (e.g., SN~1987A, Papaliolios
\etal 1989; see, however, H\"{o}flich 1990) collectively argue for asymmetry in
the explosion mechanism and/or distribution of SN ejecta.  Moreover, recent
advances in the understanding of the hydrodynamics and distribution of material
in the pre-explosion core (Bazan \& Arnett 1994; Lai \& Goldreich 2000),
coupled with results obtained through multidimensional numerical explosion
models (Burrows, Hayes, \& Fryxell 1995), imply that asphericity may be a
generic feature of the explosion process (Burrows 2000).

Sparking even more interest in SN morphology is the strong spatial and temporal
association between some ``hypernovae'' (SNe with early-time spectra
characterized by unusually broad line features; see K. Maeda's contribution to
these Proceedings) and gamma-ray bursts (GRBs; e.g., Galama \etal 1998; 
Iwamoto \etal 1998; Woosley, Eastman, \& Schmidt 1999; 
Stanek \etal 2003; Hjorth \etal 2003).
These associations have fueled the proposition that some (or, perhaps
all) core-collapse SNe explode due to the action of a ``bipolar'' jet of
material (Wheeler, Meier, \& Wilson 2002; Khokhlov \etal 1999; MacFadyen \&
Woosley 1999), as opposed to the conventional neutrino-driven mechanism
(Colgate \& White 1966; Burrows \etal 2000, and references therein).  Under
this paradigm, a GRB is only produced by those few events in which the
progenitor has lost most or all of its outer envelope material (i.e., it is a
``bare core'' collapsing), and is only observed if the jet is closely aligned
with our line of sight.  Such an explosion mechanism predicts severe
distortions from spherical symmetry in the ejecta.

\begin{figure}[ht!]
\begin{center}

 \scalebox{0.3}{
	\includegraphics{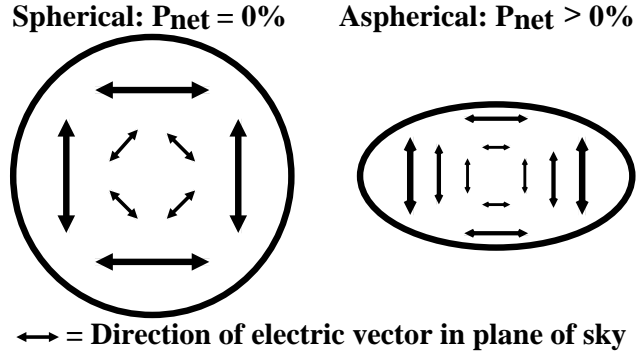}
		}
\end{center}

\caption{Polarization magnitude and direction (in the plane of the sky) for a
resolved electron-scattering atmosphere; for an unresolved source (i.e., a
supernova), only the {\it net} magnitude and direction can be measured.
\label{fig:fig1} }
\end{figure}

An exciting, emerging field is SN spectropolarimetry, an observational
technique that allows the only {\it direct} probe of early-time SN geometry.
As first pointed out by Shapiro \& Sutherland (1982; see also McCall 1984),
polarimetry of a young SN is a powerful tool for probing its geometry.  The
idea is simple: A hot young SN atmosphere is dominated by electron scattering,
which by its nature is highly polarizing.  Indeed, if we could resolve such an
atmosphere, we would measure changes in both the position angle and
strength of the polarization as a function of position in the atmosphere.  For
a spherical source that is unresolved, however, the directional components of
the electric vectors cancel exactly, yielding zero net linear polarization.  If
the source is aspherical, incomplete cancellation occurs, and a net
polarization results (Fig.~\ref{fig:fig1}).  In general, linear polarizations of
$\sim 1$\% are expected for moderate ($\sim 20$\%) SN asphericity.  The exact
polarization amount varies with the degree of asphericity, as well as with the
viewing angle and the extension and density profile of the electron-scattering
atmosphere; through comparison with theoretical models (e.g., H\"{o}flich
1991), the early-time geometry of the expanding ejecta may be derived.  In
addition to bulk asymmetry, a wealth of information about the specific nature
of the implied asphericity can be gleaned from a detailed analysis of line
features in the spectropolarimetry (e.g., Leonard \& Filippenko 2001; Leonard
\etal 2000a, 2001, 2002a; Kasen \etal 2003).

   Largely due to the difficulty of obtaining the requisite signal for all but
the brightest objects, the field of SN spectropolarimetry remained in its
infancy until quite recently.  Indeed, prior to our recent efforts and those of
a few other groups, spectropolarimetry existed only for SN 1987A in the LMC
(see Jeffery 1991 and references therein) and SN 1993J in M81 (Tran \etal
1997).  The situation has changed dramatically in the last 5 years.  Detailed
spectropolarimetric analysis now exists for more than a dozen SNe (Leonard
\etal 2000a,b, 2001, 2002a,b; Leonard \& Filippenko 2001; Howell \etal 2001;
Kasen \etal 2003; Wang \etal 2001, 2003), and the basic landscape of the young
field is becoming established.  The fundamental result is that asphericity is a
ubiquitous feature of young SNe of all types, although the nature and degree of
the asphericity vary considerably. Here we review the spectropolarimetric
characteristics of SNe, concentrating on observations made by our team at UC
Berkeley; data obtained by the Texas (Austin) group are discussed by Lifan Wang
and others in these Proceedings.

\section{Interstellar Polarization}

  It is important to first note that a difficult problem in the interpretation
of all SN polarization measurements is proper removal of interstellar
polarization (ISP), which is produced by directional extinction resulting from
aspherical dust grains along the line of sight that are aligned by some
mechanism such that their optic axes have a preferred direction.  The ISP can
contribute a large polarization to the observed signal.  Fortunately, ISP has
been well studied in the Galaxy and shown to be a smoothly varying function of
wavelength and constant with time (e.g., Serkowski, Mathewson, \& Ford 1975;
Whittet \& van Breda 1978), two properties that are not characteristics of SN
polarization.  This has allowed us to develop a number of techniques that
confidently eliminate ISP from observed SN spectropolarimetry. For example, the
Galactic component can be quantified with observations of distant,
intrinsically unpolarized Galactic ``probe stars'' along the line of sight to
the SN, while limits on the host-galaxy component can be estimated from the
observed reddening of the SN.  Alternatively, one may assume specific emission
lines or spectral regions to be intrinsically unpolarized, and derive the ISP
from the observed polarization at these wavelengths.

Improperly removed, ISP can increase or decrease the derived intrinsic
polarization, and it can change ``valleys'' into ``peaks'' (or vice versa) in
the polarization spectrum. Since it is so difficult to be certain of accurate
removal of ISP, it is generally (but not always) safest to focus on (a)
temporal changes in the polarization with multiple-epoch data, (b) distinct
line features in spectropolarimetry having high signal-to-noise ratios, and (c)
continuum polarization unlike that produced by dust. We will show near the end
of this paper, however, that technique (c) can be risky, since dust with
unusual properties may be the true cause of observed polarization.

\section{Type IIn Supernovae}

\begin{figure}[ht!]
\leavevmode
\hspace*{-0.5in}
\scalebox{0.3}{
\includegraphics{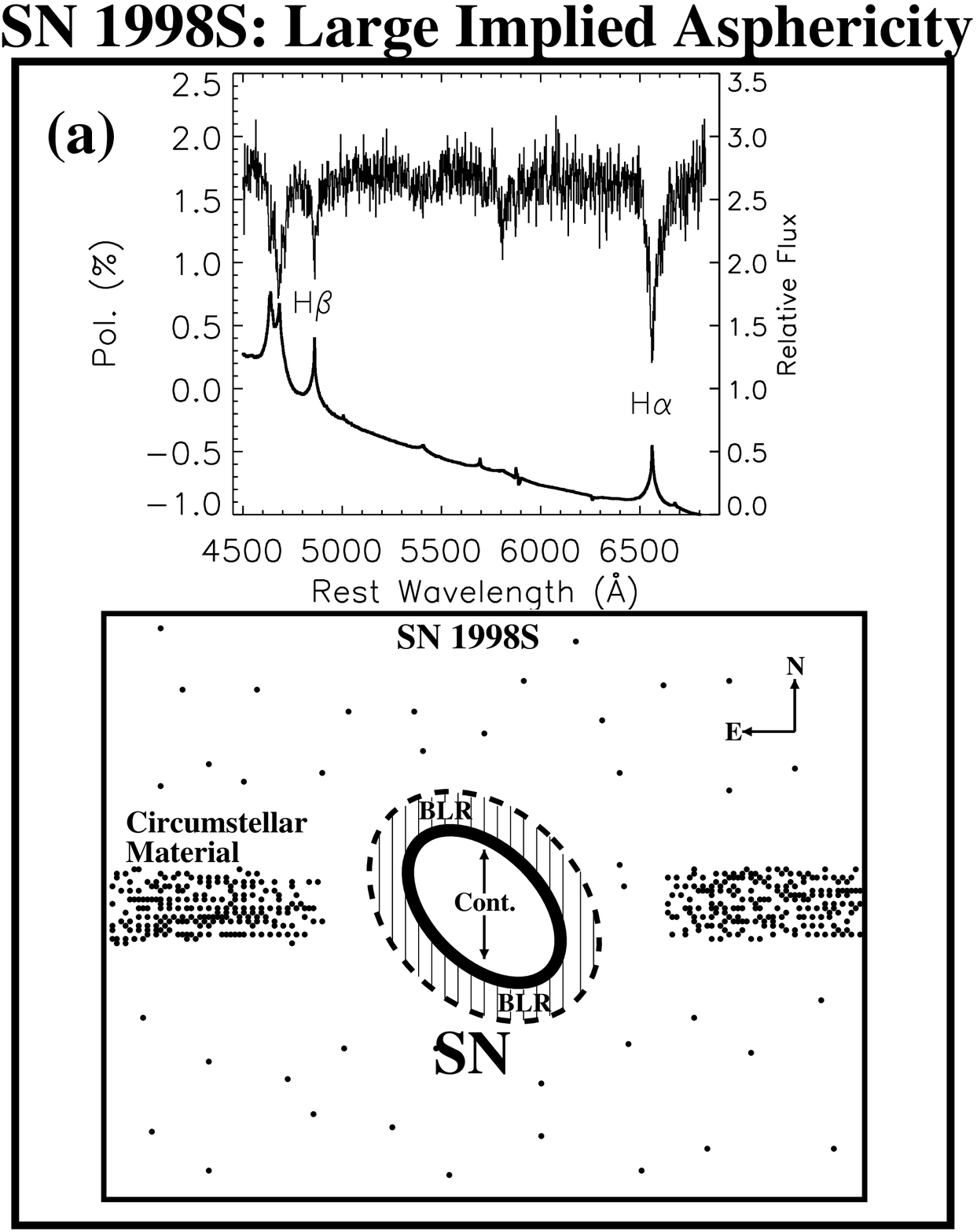}}
\scalebox{0.35}{
\hskip0.2in
\includegraphics{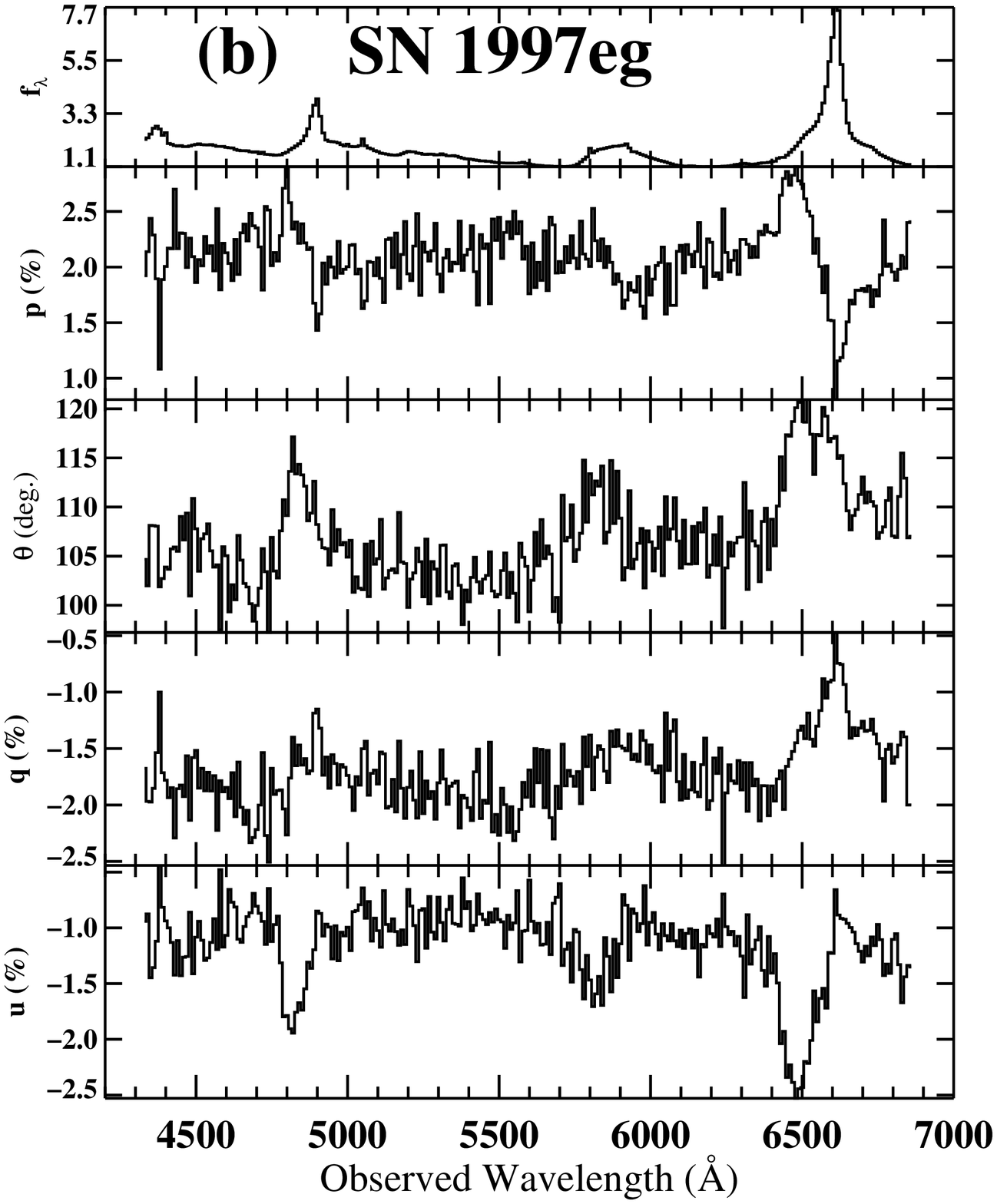}}

\caption{{\it (a)} Spectropolarimetry and flux data ({\it top, noisy} and {\it
smooth lines}, respectively), and inferred geometry ({\it bottom}) for SN
1998S, a peculiar type IIn event. {\it (b)} Polarization data for the Type IIn
SN~1997eg obtained 1998 January 17, about 55 days after discovery.  From top to
bottom are shown the total flux, in units of $10^{-15}$ ergs s$^{-1}$ cm$^{-2}$
\AA$^{-1}$, observed degree of polarization, polarization angle in the plane of
the sky, and the normalized $q$ and $u$ Stokes parameters.
\label{fig:fig2} }
\end{figure}

  We have found that Type IIn supernovae (SNe~IIn; i.e., dominated by
relatively narrow hydrogen lines --- see Filippenko 1997, and references
therein) tend to be highly polarized. One of our earliest and best-studied
examples is SN 1998S, a SN~IIn in which the progenitor star had lost most, but
not all, of its hydrogen envelope prior to exploding.  Immediately after
exploding, an intense interaction ensued between the expanding ejecta and a
dense circumstellar medium (CSM; $n_e \approx 10^7 {\rm\ cm^{-3}}$).  We
combined one early-time (3 days after discovery) Keck spectropolarimetric
observation with total-flux spectra spanning nearly 500 days to produce a
detailed study of the SN explosion and its CSM (Leonard \etal 2000a).  The high
S/N polarization spectrum is characterized by a flat continuum (at $p \approx
2$\%) with distinct changes in polarization associated with both the broad
($\gtrsi 10,000$ km s$^{-1}$) and narrow ($< 300$ km s$^{-1}$) line emission
seen in the total-flux spectrum.  The very high intrinsic polarization implies
a global asphericity of $\gtrsi 45\%$; the line profiles favor a ring-like
geometry for the circumstellar gas, generically similar to what is seen
directly in SN 1987A, but much denser and closer to the progenitor in SN~1998S
(Fig.~\ref{fig:fig2}a).  

Another SN~IIn with high polarization is SN~1997eg, for which we obtained three
epochs of spectropolarimetry over a three-month period shortly after discovery
(Nakano \& Aoki 1997). Although different in detail, the basic spectral
characteristics of this SN~IIn most closely resemble those of SN~1988Z
(Stathakis \& Sadler 1991), with narrow (unresolved, FWHM $\lesssi 200$ km
s$^{-1}$), intermediate (FWHM $\approx 2000$ km s$^{-1}$), and broad (FWHM
$\approx 15000$ km s$^{-1}$) line emission lacking the P-Cygni absorption seen
in more normal type II events dominating the spectrum.  There are clear changes
in both the magnitude (by {\bf $\sim 1$\%}) and position angle of the
polarization across strong spectral lines (e.g., ${\rm H\alpha}$;
Fig.~\ref{fig:fig2}b) in all three epochs, and the overall continuum level of
polarization changes by $\sim 1\%$ over the three months of observation.  Both
of these results argue for at least a 1\% polarization intrinsic to the SN,
although the polarization could be produced in part by an interaction of SN
ejecta with an asymmetric CSM.

\section{Type II-P Supernovae}

  It appears that SNe~II-P are not polarized much, at least shortly after the
explosion. Examples include SNe~1997ds and 1998A, which show little if any
evidence for intrinsic polarization; the weak line and continuum features are
probably partly (or mostly) due to ISP (Leonard \& Filippenko 2001). Thus, the
massive, largely intact hydrogen envelopes of the progenitors of SNe~II-P are
essentially spherical at the time of the explosion.

\begin{figure}[ht!]
\begin{center}
\vspace*{-0.6in}
\scalebox{0.40}{
\includegraphics{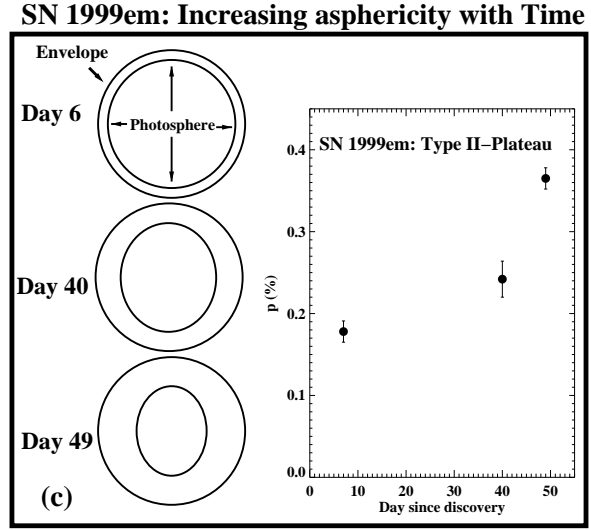}
}
\end{center}
\vspace*{-0.8in}

\caption{ The temporal increase in the polarization of the Type
II-P SN 1999em suggests greater asphericity deeper into the ejecta.
\label{fig:fig3} }
\end{figure}

   However, for SN 1999em, an extremely well-observed SN~II-P for which rare,
multi-epoch spectropolarimetry exists, we find that the polarization increased
with time (Fig.~\ref{fig:fig3}), implying a substantially spherical geometry at
early times that perhaps becomes more aspherical at late times when the deepest
layers of the ejecta are revealed. As we will see in the next sections, there
is also evidence for large polarizations in SNe that have lost much of their
envelope prior to exploding. Both lines of evidence suggest that the deeper we
peer into core-collapse events, the greater the asphericity.

The lack of evidence for early-time large-scale departures from spherical
symmetry of SNe~II-P is encouraging for the use of this class of objects as
extragalactic distance indicators through the expanding photosphere method
(EPM; Kirshner \& Kwan 1974; Eastman, Schmidt, \& Kirshner 
1996; Leonard \etal 2002c,d;
see also Leonard \etal 2003b), since this technique relies on the assumption of
a spherically symmetric flux distribution during the early stages of
development (i.e., the plateau phase).  Spectropolarimetry of SNe~II-P
therefore provides a critical test of the cosmological utility of these events,
which the small number of SNe~II-P observed thus far have passed.

\section{Other Core-Collapse Supernovae}

\subsection{SNe IIb}

   SNe~IIb are believed to be core-collapse SNe in which the progenitor lost
much of its hydrogen envelope through winds or mass transfer to a companion
star (e.g., Filippenko 1997, and references therein). The best-studied case is
SN 1993J: Tran \etal (1997) found a continuum polarization of $\sim 1$\% a
month after the explosion. More importantly, they showed that its
polarized-flux spectrum resembles the total-flux spectra of SNe~Ib, with
prominent He~I lines! The data are consistent with models in which the
polarization is produced by an asymmetric He core configuration of material.
Trammell, Hines, \& Wheeler (1993; see also H\"{o}flich 1995, H\"{o}flich \etal 1996) also found
that SN 1993J was polarized and intrinsically asymmetric, although their
derived level of interstellar polarization differed from that derived with the
more extensive data set of Tran \etal (1997).

   A subsequent SN~IIb, SN 1996cb, showed substantially similar polarization of
its spectra (Wang et al. 2001).  This was unanticipated; the polarized flux
spectra are expected to depend on the viewing angle, which should be random.
Most recently, preliminary analysis of Keck spectropolarimetry of SN~IIb 2003ed
about 1 month after explosion yields similar results: Leonard, Chornock, \&
Filippenko (2003a) find an average continuum polarization of about 1\%, with
strong modulations across the H$\alpha$ and Ca~II near-IR triplet features of
up to $\sim 1$\% in the individual Stokes parameters.  While some of the
continuum polarization may be due to interstellar dust, the changes across the
line features suggest substantial polarization due to scattering by aspherical
SN ejecta, as had been inferred for SN 1993J and SN 1996cb.

\subsection{SNe Ib and Ic}

The Texas group identified a trend, largely through broad-band polarimetry,
that the percent polarization increases along the sequence SN~II-P to IIb to Ib
to Ic (Reddy, H\"{0}flich, \& Wheeler 1999; Wheeler 2000; Wang \etal 2001).
Our group (Leonard \& Filippenko 2001; Leonard \etal 2001, 2002a,b) confirmed
this with more objects using spectropolarimetry. Certainly SNe~IIb exhibit
higher polarization than SNe~II-P, as discussed above.

\begin{figure}[ht!]
\vspace*{-0.1in}
\hspace*{-0.4in}
\scalebox{0.34}{
\includegraphics{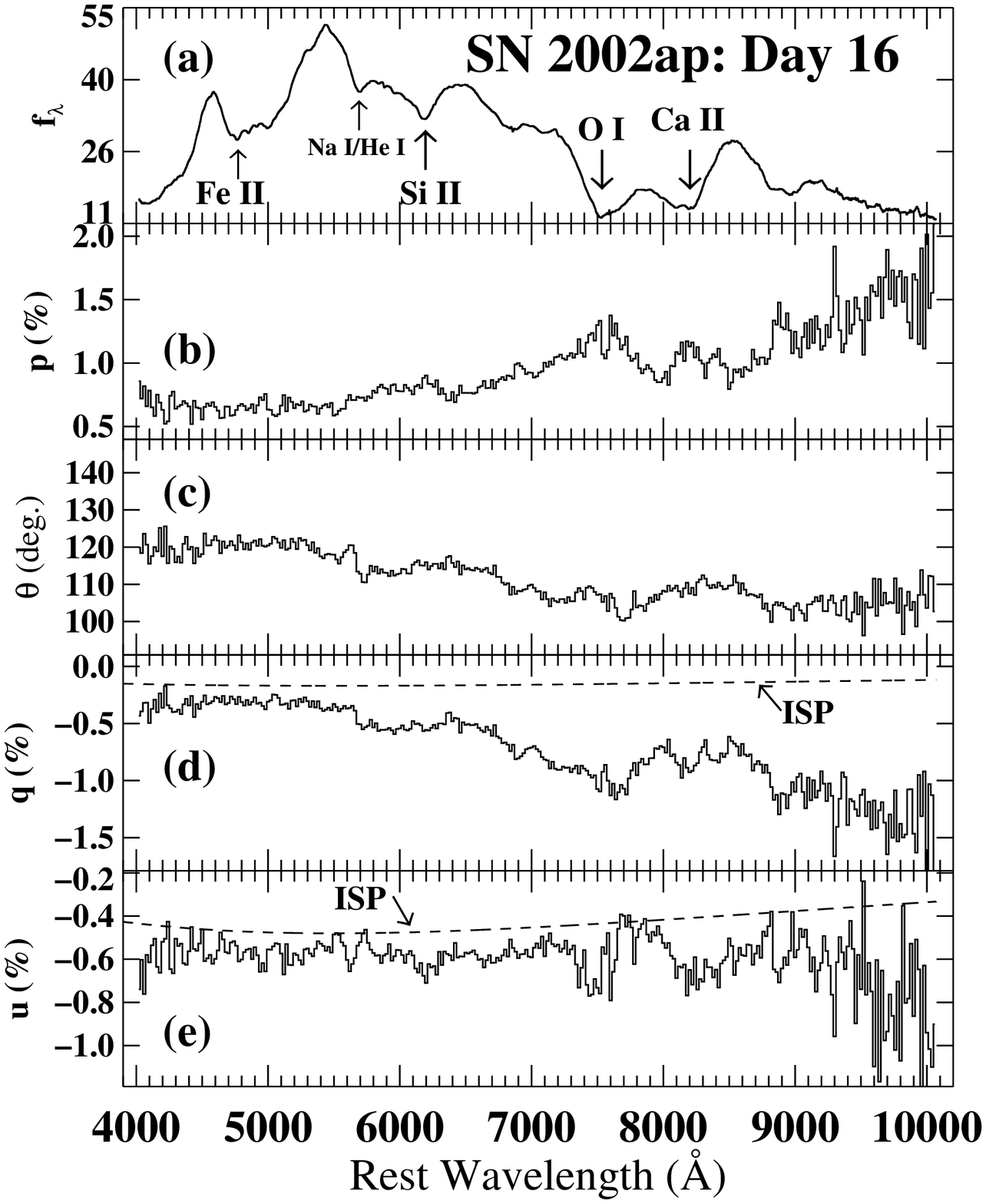}}
\scalebox{0.34}{
\hskip0.1in
\includegraphics{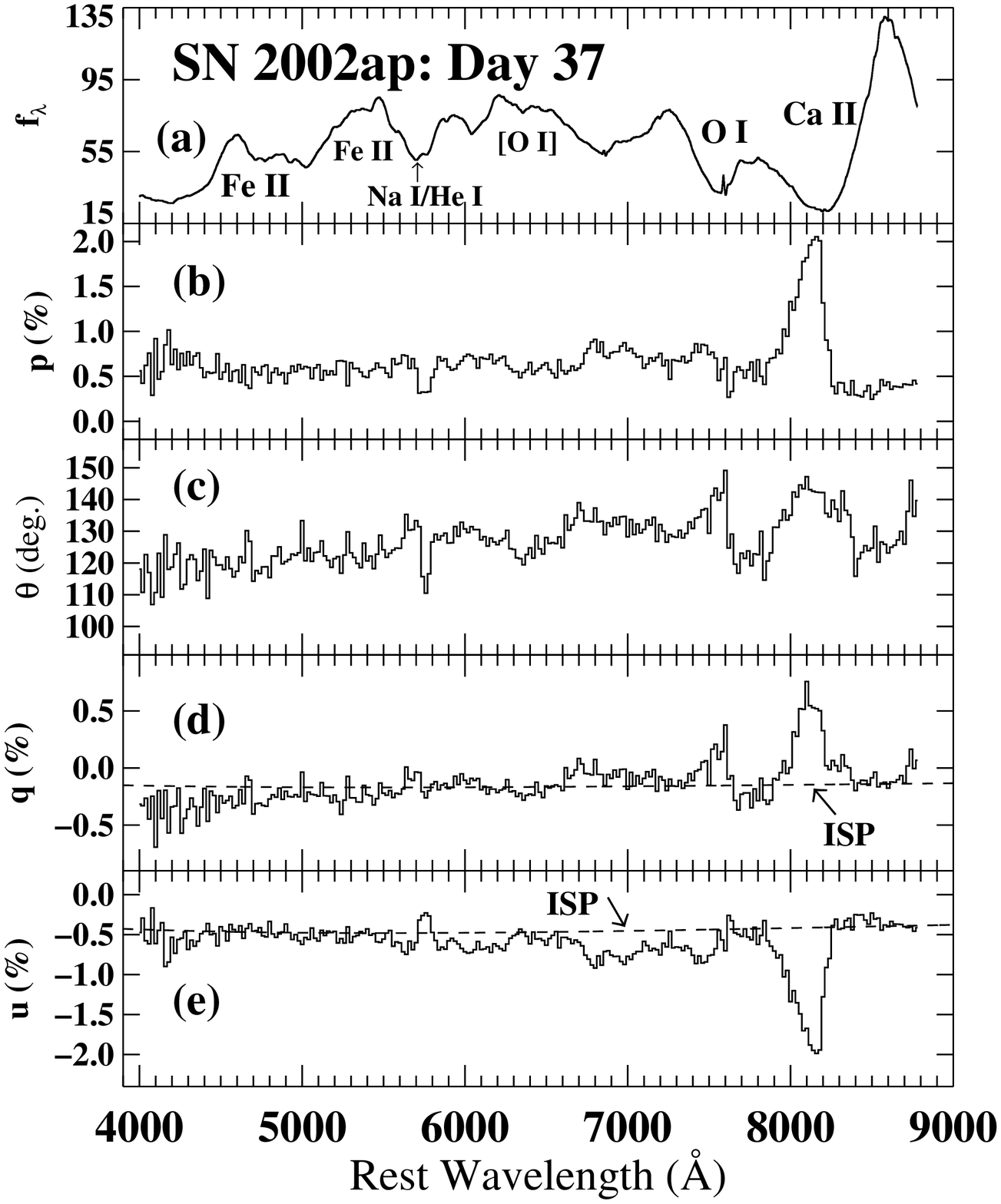}}
\vspace*{0.2in}
\caption{Polarization data (Leonard \etal 2002a) 
for SN~2002ap obtained {\it (left}) 2002 February 14
and {\it (right)} 2002 March 7, about 16 and 37 days after explosion (Mazzali
et al. 2002).  The NASA/IPAC Extragalactic Database (NED) recession velocity of
657 km s$^{-1}$ for M74 has been removed.  ({\it a}) Total flux, in units of
$10^{-15}$ ergs s$^{-1}$ cm$^{-2}$ \AA$^{-1}$, with prominent absorption
features identified by Mazzali et al. (2002) indicated.  ({\it b}) Observed
degree of polarization.  ({\it c}) Polarization angle in the plane of the
sky. ({\it d, e}) The normalized $q$ and $u$ Stokes parameters, with the level
of ISP estimated by Leonard \etal (2002a) indicated.  
\label{fig:fig4} }
\end{figure}

   A very exciting recent object is the peculiar, ``SN 1998bw-like'' SN~Ic
2002ap (Gal-Yam, Ofek, \& Shemmer 2002; Mazzali \etal 2002; Filippenko 2003;
Foley \etal 2003), which permitted analysis of the collapse of a ``bare core''
and provided several critical tests of the ``jet-induced'' SN explosion model.
Our two epochs of Keck spectropolarimetry (Leonard \etal 2002a) both show
strong evidence for intrinsic polarization, whose character changed
dramatically with time (Fig.~\ref{fig:fig4}).  Remarkably, as first noted by
Kawabata \etal (2002), the intrinsic polarized-flux optical spectrum is similar
to the total-flux optical spectrum redshifted by $0.23c$ during our first
epoch, which may indicate that much of the polarized continuum at early times
results from scattering off of electrons in a relativistic jet of material
emitted from the SN during the explosion.  Another prediction of jet-induced SN
explosion models is that intermediate-mass and heavy elements such as iron are
ejected (at high velocity) primarily along the poles whereas elements
synthesized in the progenitor (e.g., He, C, Ca, O) are preferentially located
at lower velocities near the equatorial plane in the expanding ejecta (Maeda
\etal 2002; Khokhlov \& H\"{o}flich 2001).  From a careful study of both the
temporal and spectral changes of the polarization angle, we find observational
evidence supporting a true difference in the distribution of Ca relative to
iron-group elements in the expanding ejecta (Leonard \etal 2002a).

  For core-collapse events, then, it seems that the closer we probe to the
heart of the explosion, the greater the polarization and, hence, the asymmetry.
The small, but temporally increasing polarization of SNe II-P coupled with the
high polarization of stripped-envelope SNe implicate an explosion mechanism
that is highly asymmetric for core-collapse events.  The current speculation is
that the presence of a thick hydrogen envelope dampens the observed asymmetry.
Thus, explosion asymmetry, or asymmetry in the collapsing Chandrasekhar core,
may play a dominant role in the explanation of pulsar velocities, the mixing of
radioactive material seen far out into the ejecta of young SNe (e.g., SN~1987A;
Sunyaev \etal 1987), and even GRBs.

\section{Type Ia Supernovae}

   Although the nearly unanimous consensus is that a mass-accreting white dwarf
approaching the Chandrasekhar limit is the progenitor of SNe~Ia, observational
evidence for this scenario remains elusive.  Given a progenitor that is
actively accreting material at the time of the explosion, and is likely to be
part of a binary system, it seems probable that some distortion of the ejecta
should exist, and lend itself to detection through spectropolarimetry. Evidence
for intrinsic polarization in SNe~Ia has been more difficult to find than for
the core-collapse events, and initial investigations found only minimal
evidence for polarization, $p \lesssi 0.2$\% (e.g., Wang, Wheeler, \&
H\"{o}flich 1997; Wheeler 2000).

In the last few years, though, significant advances have been made. Leonard,
Filippenko, \& Matheson (2000b) reported the first convincing, albeit weak,
features in the polarization of an SN~Ia, SN 1997dt.  The most thorough
polarization study of an SN~Ia was conducted by Howell \etal (2001): the
subluminous SN~Ia 1999by exhibits a change in $p$ by $\sim 0.8$\% from
4800~\AA\ to 7100~\AA, consistent with an asphericity of about 20\% observed
equator-on. However, the unusual (subluminous) nature of SN 1999by still left
some doubt about intrinsic polarization in {\it normal} SNe~Ia.

   That doubt has recently been put to rest, with the work of Wang \etal (2003)
and Kasen \etal (2003) on SN 2001el. Similar to SN 1999by, after subtraction of
the ISP, the percent polarization increases from blue to red wavelengths in
spectra obtained 1 week before maximum brightness.  However, the extraordinary
feature here is the existence of distinct high-velocity Ca~II near-IR triplet
absorption ($v$ = 18,000--25,000 km s$^{-1}$) in addition to the ``normal"
lower-velocity Ca~II feature.  A similar, but much weaker such feature had been
previously observed in SN 1994D, and perhaps in other SNe~Ia as well; the
number of pre-maximum spectra covering the near-IR spectral range is small.
The polarization is seen to increase dramatically in this feature, which Kasen
\etal (2003) interpret in terms of a detached clumpy shell of high-velocity
material that partially obscures the underlying photosphere and causes an
incomplete cancellation of the polarization of the photospheric light, 
giving rise to the polarization peak. From a detailed study of the continuum
polarization of $\sim$0.2\% in SN 2001el, Wang \etal (2003) conclude that the
continuum-producing region itself likely has an axis ratio of about 0.9.

\begin{figure}[ht!]
\begin{center}
\rotatebox{90}{
\scalebox{0.35}{
\includegraphics{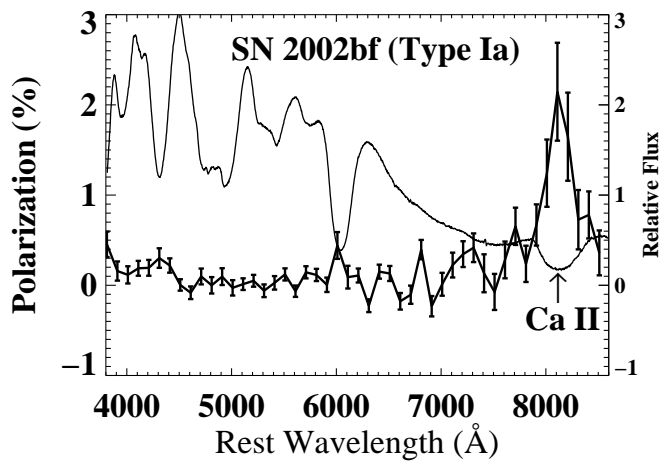}
}}
\end{center}

\caption{Flux ({\it thin smooth curve}) and observed polarization 
({\it binned points with error bars}) of the Type Ia
SN~2002bf.  ISP is not a major concern for this SN, since $E(B-V) = 0.01$ mag
for the Milky Way (Schlegel, Finkbeiner, \& Davis 1998), and there is no
Na~I~D absorption from the host galaxy.
\label{fig:fig5} }
\end{figure}

Our group obtained maximum-light spectropolarimetry of the SN~Ia 2002bf using
the Keck-I telescope.  This supernova is characterized by uncharacteristically
large photospheric velocities (Filippenko \etal 2002), with no obviously
detached high-velocity components, although it remains to be seen if it was in
any other way peculiar.  Figure~\ref{fig:fig5} shows the total-flux spectrum and
the percent polarization near maximum light. There is a large increase in
polarization in the photospheric Ca~II near-IR trough, in which the
polarization increases from almost zero to about 2\%, which is actually similar
behavior to what was seen in the SN~Ic 2002ap discussed earlier
(Fig.~\ref{fig:fig4}). When interpreted in terms of the simple geometric dilution
model used for SN 2002ap, this polarization increase would imply a global
asphericity of at least 15\%; however, as we have seen from Kasen \etal (2003),
other interpretations are possible that might not need to resort to global
asphericity, including clumpy ejecta.  On one of our latest Keck runs, in May
2003, we observed the SN~Ia 2003du, and preliminary reductions of the
spectropolarimetry show it to be similar to SN 2002bf.  The continuum
polarization slowly rises from blue to red, increasing from 0\% to about 0.2\%,
and there are modulations across both the Si~II 6350 and Ca~II near-IR
features.

  Thus, from these studies, we can conclusively say that at least some SNe~Ia
are intrinsically polarized.  However, this field is still very much in its
infancy, and more work, especially of the theoretical sort, needs to be done to
gain confidence in the interpretations of the growing database of empirical
data. We have not yet seen clear evidence for binarity of the SN~Ia
progenitors, which is puzzling.

\section{SN Spectropolarimetry as a Probe of Interstellar Dust} 

\begin{figure}[ht!]
\begin{center}
\scalebox{0.35}{
\includegraphics{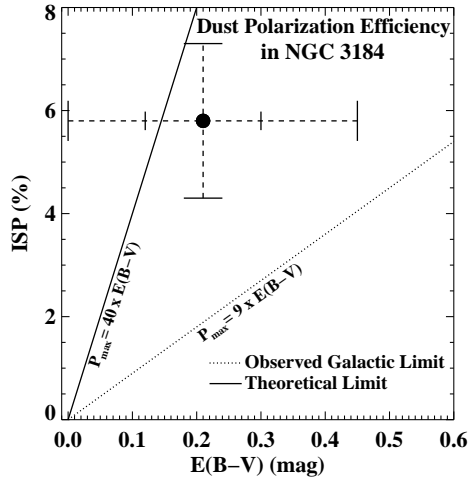}
}
\end{center}

\caption{ISP and reddening along the line-of-sight to SN~1999gi in NGC~3184, with
the observed polarization and best reddening estimate indicated by the {\it
filled circle}.  The $1\ \sigma$ reddening uncertainty is
indicated ({\it short error bars}), as are the lower and upper limits for both
the reddening and ISP ({\it long error bars}).  Also shown is the empirical
limit on the polarization efficiency of MW dust ({\it dotted line}; Serkowski
et al. 1975), as well as the theoretical upper limit for dust grains consisting
of completely aligned infinite dielectric cylinders ({\it solid line}; Whittet
\etal 1992).
\label{fig:fig6} }
\end{figure}

A rich byproduct of supernova spectropolarimetry is the ability to study
fundamental properties of the interstellar dust in external galaxies, a
research area with notoriously few direct observational diagnostics.  The
degree of polarization produced for a given amount of extinction (or reddening)
is referred to as the ``polarization efficiency'' of the intervening dust
grains.  Through study of the polarization of thousands of Galactic stars, an
upper bound on the polarization efficiency of Galactic dust has been derived
(Serkowski et al. 1975): ${\rm ISP} / E_{B-V} < 9.0\%\ {\rm mag}^{-1}$.  Our
observations of the Type~II-P SN~1999gi (Leonard \& Filippenko 2001) resulted
in the discovery of an extraordinarily high polarization efficiency for the
dust along the line-of-sight in the host galaxy, NGC~3184: ISP$/E(B-V)=
31^{+22}_{-9}\% {\rm\ mag}^{-1}$ (Leonard \etal 2002b).  This is more than
three times the empirical Galactic limit, strains the theoretical Mie
scattering limit (${\rm ISP}/E(B-V) < 40\% {\rm\ mag}^{-1}$; Whittet \etal
1992), and represents the highest polarization efficiency yet confirmed for a
single sight line in either the Milky Way or an external galaxy
(Fig.~\ref{fig:fig6}).  While the polarization properties of the dust grains along
the line-of-sight in NGC~3184 are quite unusual, our analysis also revealed the
average size of the grains to be quite similar to the inferred size of typical
dust grains in the Milky Way, $\sim0.14\ \mu{\rm m}$.  We speculate that the
very high polarization efficiency of the grains may indicate an unusually
regular magnetic field in NGC~3184 or even a different dust grain alignment
mechanism than has traditionally been assumed.

\vspace{0.3cm}
{\bf Acknowledgments}.

We thank Craig Wheeler, in whose honor this meeting was held, for many
stimulating discussions of supernovae and other (mostly explosive) phenomena
over the years. A.V.F. is grateful to the workshop organizers for travel
support and for their incredible patience while waiting for the written version
of his presentation.  The research of A.V.F.'s group at UC Berkeley is
supported by NSF grant AST-0307894, as well as by NASA through grants GO-9155
and GO-9428 from the Space Telescope Science Institute, which is operated by
the Association of Universities for Research in Astronomy, Inc., under NASA
contract NAS~5-26555.  D.C.L. acknowledges additional support by NASA through
the American Astronomical Society's Small Research Grant Program.  Most of our
spectropolarimetry was obtained at the W. M. Keck Observatory, which is
operated as a scientific partnership among the California Institute of
Technology, the University of California, and NASA; the observatory was made
possible by the generous financial support of the W. M. Keck Foundation.

\end{document}